\documentclass[preprint,showpacs,preprintnumbers,amsmath,amssymb]{revtex4}

\def\undersim#1{\setbox9\hbox{${#1}$}{#1}\kern-\wd9\lower
    2.5pt \hbox{\lower\dp9\hbox to \wd9{\hss $_\sim$\hss}}}
\usepackage{amssymb}
\usepackage{amsmath}
\usepackage{graphicx}

\def\undersim#1{\setbox9\hbox{${#1}$}{#1}\kern-\wd9\lower
    2.5pt \hbox{\lower\dp9\hbox to \wd9{\hss $_\sim$\hss}}}

\def\mv{{\mathbf v}}
\def\mr{{\mathbf r}}

\def\mr{{\mathbf r}}

\def\mk{{\mathbf k}}

\begin{document}

\title{Squeezed correlations of $\phi$ meson pairs for hydrodynamic
sources in high-energy heavy-ion collisions}

\author{Yong Zhang$^1$}
\author{Jing Yang$^1$}
\author{Wei-Ning Zhang$^{1,\,2,}$\footnote{wnzhang@dlut.edu.cn}}
\affiliation{$^1$School of Physics and Optoelectronic Technology,
Dalian University of Technology, Dalian, Liaoning 116024, China\\
$^2$Department of Physics, Harbin Institute of Technology, Harbin,
Heilongjiang 150006, China}


\begin{abstract}
In the hot and dense hadronic sources formed in high-energy heavy-ion
collisions, the particle interactions in medium might lead to a squeezed
back-to-back correlation (BBC) of boson-antiboson pairs.  We calculate
the BBC functions of $\phi\phi$ for sources evolving hydrodynamically
in ($2+1$) dimensions and with longitudinal boost invariance.
The BBC functions for hydrodynamic sources exhibit oscillations as
a function of the particle momentum because the temporal distributions of
hydrodynamic sources have sharp falls to 0 at large evolving times.
The dependences of the BBC functions on the directions of the particle
momentum are investigated.  For transverse anisotropic sources, the
BBC functions are minimum when the azimuthal angles of the particles
reach 0.  The BBC functions increase with decreasing absolute value
of the particle pseudorapidity.  The oscillations and the dependences
on the particle azimuthal angle and pseudorapidity are the significant
signatures for detecting the BBC in high-energy heavy-ion collisions.


\end{abstract}

\pacs{25.75.Gz, 25.75.Ld, 21.65.jk}
\maketitle

\section{Introduction}
In the hot and dense hadronic sources formed in high energy heavy ion collisions,
the mass modification of particles in medium can lead to a squeezed back-to-back
correlation (BBC) of the boson-antiboson pair \cite{AsaCso96,AsaCsoGyu99}.
This BBC is the result of a quantum mechanical transformation relating in-medium
quasiparticles to the two-mode squeezed states of their free observable counterparts,
through a Bogoliubov transformation between the creation (annihilation) operators
of the quasiparticles and the free observable particles
\cite{AsaCso96,AsaCsoGyu99,Padula06}.
The investigations of the BBC of boson-antiboson pairs may provide a new way for
people to understand the thermal and dynamical properties of the hadronic sources
in high energy heavy ion collisions.

Denote $a_\mk\, (a^\dagger_\mk)$ the annihilation (creation) operator of the
free boson with momentum $\mk$ and mass $m$, and $b_\mk\, (b^\dagger_\mk)$ the 
annihilation (creation) operator of the corresponding quasiparticle with momentum
$\mk$ and modified mass $m_{\!*}$ in homogeneous medium; they are related by
the Bogoliubov transformation \cite{AsaCso96,AsaCsoGyu99}
\begin{equation}
a_{\mk} = c_{\mk}\,b_{\mk} + s^*_{-\mk}\,b^\dagger_{-\mk},
\vspace*{-2mm}
\end{equation}
where
\begin{equation}
c_{\mk} = \cosh f_{\mk}\,,~~~~~s_{\mk} = \sinh f_{\mk}\,,~~~~~
f_{\mk} = \frac{1}{2} \log (\omega_{\mk}/\Omega_{\mk}),
\end{equation}
\begin{equation}
\omega_\mk=\sqrt{\mk^2 + m^2}\,,~~~~~\Omega_\mk=\sqrt{\mk^2+m_{\!*}^2}\,.
\end{equation}
The BBC function is defined as \cite{AsaCso96,AsaCsoGyu99}
\begin{equation}
\label{BBCf}
C(\mk,-\mk) = 1 + \frac{|G_s(\mk,-\mk)|^2}{G_c(\mk,\mk) G_c(-\mk,-\mk)},
\end{equation}
where $G_c(\mk_1,\mk_2)$ and $G_s(\mk_1,\mk_2)$ are the chaotic and squeezed
amplitudes, respectively,
\begin{equation}
G_c(\mk_1,\mk_2)=\sqrt{\omega_{\mk_1}\omega_{\mk_2}}\,\langle
a^\dagger_{\mk_1} a_{\mk_2} \rangle,
\end{equation}
\begin{equation}
G_s(\mk_1,\mk_2)=\sqrt{\omega_{\mk_1}\omega_{\mk_2} }\,\langle
a_{\mk_1} a_{\mk_2}\rangle,
\end{equation}
where $\langle \cdots \rangle$ indicates the ensemble average.  The BBC function 
for a homogeneous source with volume $V$ and temperature $T$ can be expressed
as
\cite{AsaCsoGyu99}
\begin{equation}
\label{BBCf1}
C(\mk,-\mk) = 1 + \frac{V\,|c_{\mk}\,s_{\mk}^*\,n_{\mk} +c_{-\mk}\,s_{-\mk}^*
\,(n_{-\mk}+1)|^2}{V\,[n_1(\mk)\,n_1(-\mk)]},
\end{equation}
where
\begin{equation}
n_{\mk}=\frac{1}{\exp(\Omega_{\mk}/T)-1},
\end{equation}
\begin{equation}
n_1(\mk)=|c_{\mk}|^2\,n_{\mk}+|s_{-\mk}|^2(n_{-\mk}+1).
\end{equation}

In Ref. \cite{Padula06}, S. Padula {\it et al.} put forward the formulism 
of the BBC function for the local-equilibrium evolving system and studied 
the BBC functions of $\phi\phi$ for expanding sources with a Gaussian space
profile.  Recently, the BBC functions of $K^+ K^-$ were investigated
\cite{Padula10} for expanding Gaussian sources, and a method was suggested
\cite{Padula10a} to search for the squeezed BBC in heavy-ion collisions
at the Relativistic Heavy Ion Collider (RHIC) and the Large Hadron Collider
(LHC).  In Refs. \cite{YZHANG14} and \cite{YZHANG15}, we calculated the BBC 
functions of relativistic $\phi\phi$ and $K^+K^-$ pairs for spherical and 
ellipsoid expanding Gaussian sources.  The relativistic effect on the BBC 
functions \cite{YZHANG14} and the dependence of the BBC functions on the 
direction of the particle momentum for the anisotropic sources \cite{YZHANG15} 
are investigated.  However, all the source density distributions used in the
calculations in previous works are space-time separated, having Gaussian 
spatial distributions and an independent temporal distribution of exponential 
decay or others \cite{Padula06,Padula10,Padula10a,YZHANG14,YZHANG15}.  
Investigations of the BBC based on more realistic space-time evolving
source models will be of interest in high-energy heavy-ion collisions.

Relativistic hydrodynamics has been extensively applied to high-energy 
heavy-ion collisions.  In this work, we use the ideal relativistic 
hydrodynamics in $2+1$ dimensions to describe the transverse expansion of 
sources with zero net baryon density and combine the Bjorken boost-invariant 
hypothesis \cite{Bjo83} for the source longitudinal evolution.  As a first 
step to the study of the BBC for a more realistic source model, these 
descriptions are suitable for the heavy-ion collisions at the RHIC top 
energy and the LHC energy \cite{{Ris98,KolHei03,Bay83,Gyu97,Ris9596,BLM96,
BLM04,BLM08,BLM11,Kol00,KolRap03,She10,ZhaEfa2}}.
We investigate the BBC functions of $\phi$ meson pairs for the hydrodynamic
sources.  The results indicate that the BBC functions of $\phi\phi$ exhibit
oscillations as a function of the particle momentum and vary with the particle
azimuthal angle and pseudorapidity.  The oscillations and the dependences on
particle azimuthal angle and pseudorapidity are the significant signatures
for detecting the BBC in high-energy heavy-ion collisions.

The rest of this paper is organized as follows.  In Sec. II, we present
the calculation formulas of the BBC function for hydrodynamic sources.
In Sec. III, we investigate the BBC functions of $\phi\phi$ for 
hydrodynamic sources with different initial geometries and energy densities.
The oscillations of BBC functions as a function of the particle momentum
and the dependences of BBC functions on the directions of the particle
momentum are also discussed in this section.  Finally, a summary and 
conclusions of this paper are given in Sec. IV.

\section{Calculations of BBC functions for hydrodynamic sources}
The description of ideal hydrodynamics for the system with zero net-baryon
density is defined by the local conservations of energy and momentum
\cite{Ris98,KolHei03},
\begin{equation}\label{hy1}
\partial_{\mu}T^{\mu\nu}(r)=0,
\end{equation}
where $T^{\mu\nu}(r)\!=\left[\epsilon(r)\!+\!{\cal P}(r)\right]u^{\mu}(r)
u^{\nu}(r)-{\cal P}(r)g^{\mu\nu}$ is the density tensor of energy-momentum
ideal fluid at space-time coordinate $r$, $\epsilon(r)$ and ${\cal P}(r)$
are the energy density and pressure in the local rest frame of the fluid
element at $r$, which move at velocity $\textbf{\emph{v}}(r)$, $u^{\mu}=\gamma(1,\textbf{\emph{v}})$ is the four-velocity, $\gamma\!=\!(1\! -\!\textbf{\emph{v}}^{2})^{-1/2}$, and $g^{\mu\nu}\!=\mathrm{diag}(+,-,-,-)$
is the Minkowski metric tensor.  Under the assumption of Bjorken longitudinal
boost invariance \cite{Bjo83}, we need only to solve the transverse equations
of motion in the $z=0$ plane, and the hydrodynamic solutions at $z\ne0~(v^z=z/t)$
can be obtained by the longitudinal boost invariance hypothesis
\cite{Bay83,Gyu97}.

From Eq. (\ref{hy1}) we have the transverse equations in the $z=0$ plane,
\begin{eqnarray}\label{hyeq2}
&&\hspace*{-8mm}\partial_t {\cal E}+\partial_x [({\cal E}\!+\!{\cal P})v^x]
+\partial_y [({\cal E}\!+\!{\cal P}) v^y]=-{\cal F}({\cal E},{\cal P},t),
\nonumber\\
&&\hspace*{-8mm}\partial_t {\cal M}^x+\partial_x({\cal M}^x v^x\!+\!{\cal P})
+\partial_y ({\cal M}^x v^y)=-{\cal G}({\cal M}^x,t),\\
&&\hspace*{-8mm}\partial_t {\cal M}^y+\partial_x({\cal M}^y v^x)+\partial_y
({\cal M}^y v^y\!+\!{\cal P})=-{\cal G}({\cal M}^y,t),
\nonumber
\end{eqnarray}
where ${\cal E}=T^{00}=\gamma^2(\epsilon+{\cal P})-{\cal P} $, ${\cal M}^i=
T^{0i}=\gamma^2(\epsilon+{\cal P})v^i,\,(i=x,y)$, ${\cal F}({\cal E},
{\cal P},\,t)=({\cal E}+{\cal P})/t$, and ${\cal G}({\cal M}^i,t)={\cal M}^i
\!/t$.  In equation set (\ref{hyeq2}) there are $\epsilon$, ${\cal P}$, $v^x$,
and $v^y$ four variables.  So an equation of state, ${\cal P}(\epsilon)$,
is needed to enclose the equation set.  In the calculations, we use the 
equation of state of s95p-PCE, which combines the hadron resonance gas at 
low temperatures and the lattice QCD results at high temperatures \cite{She10}.  
We assume that the system reaches the static local equilibrium at $\tau_0=0.6$ 
fm/$c$ after the collision, and take the initial energy density distribution 
in the transverse plane as the Gaussian distribution,
\begin{equation}
\epsilon = \epsilon_0\exp[-x^2/(2R_x^2)-y^2/(2R_y^2)],
\end{equation}
where $\epsilon_0$ and $R_i~(i=x,y)$ are the parameters of the initial source
energy density and radii.  With the equation of state and the initial energy 
density we can solve equation set (\ref{hyeq2}) using the relativistic HLLE 
scheme and Sod's operation splitting method
\cite{HLLE,Ris98,Ris9596,ZhaEfa,ZhaEfa2,Zha04,Yu08Yin12,Sod77}.

For hydrodynamic sources, with the formula derived by Makhlin and Sinyukov
\cite{MakhSiny}, the chaotic and squeezed amplitudes can be expressed as \cite{AsaCsoGyu99,Padula06}
\begin{eqnarray}
\label{Gchydro}
&& G_c({\mk_1},{\mk_2})\!=\!\int \frac{d^4\sigma_{\mu}(r)}{(2\pi)^3}
K^\mu_{1,2}\, e^{i\,q_{1,2}\cdot r}\,\! \Bigl\{|c'_{\mk'_1,\mk'_2}|^2\,
n'_{\mk'_1,\mk'_2}~~~~~~\nonumber \\
&& \hspace*{19mm}
+\,|s'_{-\mk'_1,-\mk'_2}|^2\,[\,n'_{-\mk'_1,-\mk'_2}+1]\Bigr\},
\end{eqnarray}
\begin{eqnarray}
\label{Gshydro}
&& G_s({\mk_1},{\mk_2})\!=\!\int \frac{d^4\sigma_{\mu}(r)}{(2\pi)^3}
K^\mu_{1,2}\, e^{2 i\,K_{1,2}\cdot r}\!\Bigl\{s'^*_{-\mk'_1,\mk'_2}
c'_{\mk'_2,-\mk'_1}~~~~~\nonumber \\
&& \hspace*{18mm}
\times n'_{-\mk'_1,\mk'_2}+c'_{\mk'_1,-\mk'_2} s'^*_{-\mk'_2,\mk'_1}
[n'_{\mk'_1,-\mk'_2} + 1] \Bigr\}.
\end{eqnarray}
Here $d^4\sigma_{\mu}(r)$ is the four-dimension element of freeze-out
hypersurface, $q^{\mu}_{1,2}=k^{\mu}_1-k^{\mu}_2$, $K^{\mu}_{1,2}=
(k^{\mu}_1+k^{\mu}_2)/2$, and $\mk_i'$ is the local-frame momentum
corresponding to $\mk_i~(i=1,2)$.  The other local variables are:
\begin{equation}
c'_{\pm\mk'_1,\pm\mk'_2}=\cosh[\,f'_{\pm\mk'_1,\pm\mk'_2}\,],
\end{equation}
\begin{equation}
s'_{\pm\mk'_1,\pm\mk'_2}=\sinh[\,f'_{\pm\mk'_1,\pm\mk'_2}\,],
\end{equation}
\begin{eqnarray}
&&\hspace*{-5mm}f'_{\pm\mk'_1,\pm\mk'_2}=\frac{1}{2} \log \left[(
\omega'_{\mk'_1}+\omega'_{\mk'_2})/(\Omega'_{\mk'_1}+\Omega'_{\mk'_2})
\right]\nonumber\\
&&\hspace*{10mm}=\frac{1}{2} \log \left[K^{\mu}_{1,2}u_{\mu}(r)/
K^{*\nu}_{1,2}u_{\nu}(r)\right]\nonumber\\
&&\hspace*{10mm}\equiv f_{\mk_1,\,\mk_2}(r),
\end{eqnarray}
\begin{eqnarray}
&&\hspace*{-7mm}\omega'_{\mk'_i}(r)=\sqrt{\mk'^2_i(r)+m^2}=
k^{\mu}_i u_{\mu}(r)\nonumber\\
&&\hspace*{5mm}=\gamma_\mv\,[\,\omega_{\mk_i}-\mk_i \cdot \mv(r)\,],
\end{eqnarray}
\begin{eqnarray}
\label{Omp}
&&\hspace*{-7mm}\Omega'_{\mk'_i}(r)=\sqrt{\mk'^2_i(r)+m_*^2}
\nonumber\\
&&\hspace*{5mm}=\sqrt{[k^{\mu}_i u_{\mu}(r)]^2-m^2+m_*^2}\nonumber\\
&&\hspace*{5mm}=k^{*\mu}_i u_{\mu}(r),
\end{eqnarray}
\begin{eqnarray}
\label{nkk}
&&\hspace*{-8mm}n'_{\pm\mk'_1,\pm\mk'_2}=\exp\left\{-\left[\frac{1}{2}
\Big (\Omega'_{\mk'_1}+\Omega'_{\mk'_2}\Big)-\mu_{1,2}(r)\right]
\Big/ T(r)\right\}\nonumber\\
&&\hspace*{8mm}=\exp{\{-[K^{*\mu}_{1,2} u_\mu(r)-\mu_{1,2}(r)]
\,/\,T(r)\}}\nonumber\\
&&\hspace*{8mm}\equiv n_{\mk_1,\,\mk_2}(r),
\end{eqnarray}
where, $K^{*\mu}_{1,2}=(k^{*\mu}_1+k^{*\mu}_2)/2$ is the pair four-momenta
of the quasiparticles in medium, and $u^{\mu}(r)=\gamma_\mv[1,\mv(r)]$, 
$\mu_{1,2}(r)$, and $T(r)$ are the source four-velocity, the pair
chemical potential, and the source temperature at particle freeze-out,
respectively.  Equation (\ref{Omp}) gives the relationship between $k^{*\mu}
u_{\mu}(r)$ and $k^{\mu}u_{\mu}(r)$, which is used in calculating
$f_{\mk1,\mk2} (r)$ and $n_{\mk1,\mk2}(r)$.

\section{BBC results for hydrodynamic sources}
\subsection{Source distributions}
For hydrodynamic sources with a Bjorken cylinder, the four-dimension
element of the freeze-out hypersurface can be written as
\begin{equation}
d^4\sigma_{\mu}(r)=f_{\mu}(\tau, \mr_{\perp}, \eta)\, d\tau d^2
\mr_{\perp} d\eta,
\end{equation}
where $\tau$, $\mr_{\perp}$, and $\eta$ are the proper time, transverse
coordinate, and space-time rapidity of the element.  The function $f_{\mu}
(\tau,\mr_{\perp},\eta)$ is related to the freeze-out mechanism that is
considered, and $K^{\mu}_{1,2}f_{\mu}(\tau,\mr_{\perp},\eta)$ corresponds
to the source distributions of proper time and space in the calculations
[see Eqs. (\ref{Gchydro}) and (\ref{Gshydro})].  In this work we assume
that $\phi$ mesons are frozen out at a fixed temperature $T_f$ and use the
AZHYDRO technique \cite{Kol00,KolRap03,KolHei03} to calculate the freeze-out
hypersurface element.

\begin{figure}[htbp]
\includegraphics[scale=0.40]{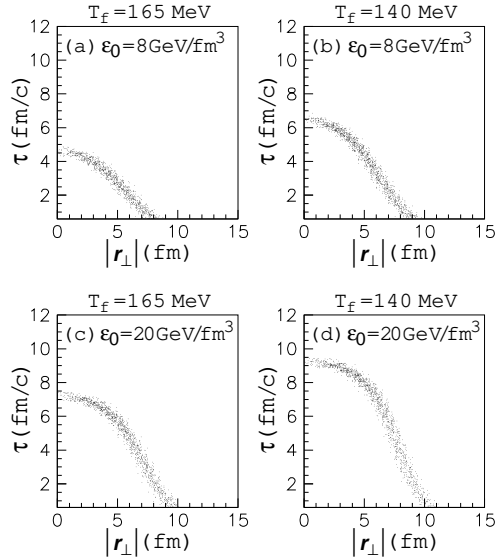}
\vspace*{-5mm}
\caption{Distributions of the freeze-out points of $\phi$ mesons in the 
$z=0$ plane for the initial conditions $\epsilon_0=$ 8 and 20 GeV/fm$^3$ 
and $R_x=R_y=4$ fm. }
\label{disfo}
\end{figure}

\begin{figure}[htbp]
\vspace*{5mm}
\includegraphics[scale=0.48]{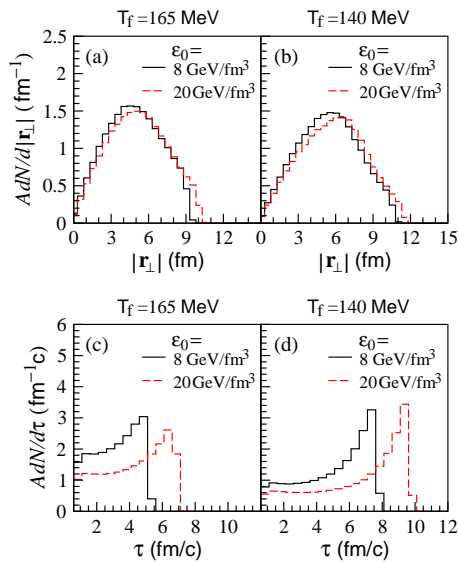}
\vspace*{-3mm}
\caption{(Color online) Normalized distributions of transverse coordinate
and time of the $\phi$ freeze-out points in the $z=0$ plane for the same 
initial conditions as in Fig. \ref{disfo}. }
\label{desrt}
\end{figure}

We show in Fig. \ref{disfo} the distributions of the freeze-out points
(source points) of $\phi$ mesons in the $z=0$ plane for the initial conditions
$\epsilon_0=$ 8 and 20 GeV/fm$^3$ and $R_x=R_y=4$ fm.  The distribution profile
for the lower $T_f$ is wider than that for the higher $T_f$ because of the
source expansion.  And the distribution profiles increase with increasing
initial energy density.  In Fig. \ref{desrt}, we show the normalized
distributions of the transverse coordinate and time of the $\phi$ freeze-out
points in the $z=0$ plane, which are obtained by projecting the two-dimensional 
distributions in Fig. \ref{disfo} to the coordinate and time axes,
respectively.  One can see that the transverse-coordinate distributions
are similar to Gaussian distributions.  The temporal distributions increase
with time nonlinearly and have sharp falls to 0 at long evolving times.
The widths of the spatial and temporal distributions increase with increasing
initial energy density and decrease with increasing freeze-out temperature.

\subsection{BBC functions}

\begin{figure}[htbp]
\includegraphics[scale=0.65]{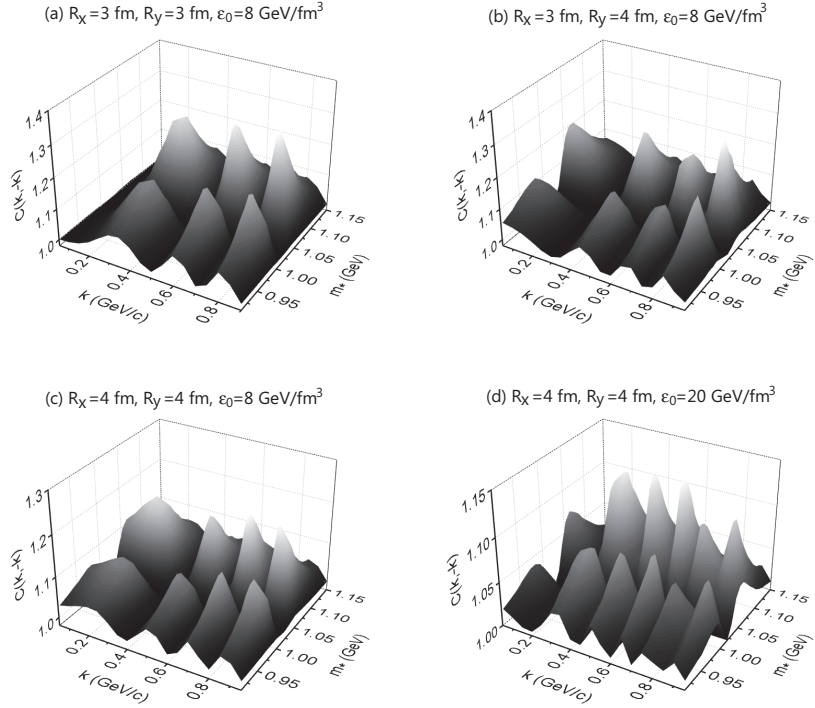}
\vspace*{-5mm}
\caption{BBC functions of $\phi\phi$ for the hydrodynamic sources
with different initial conditions and $T_f=140$ MeV. }
\label{BBCFP34}
\end{figure}

In Fig. \ref{BBCFP34},  we plot the BBC functions of $\phi\phi$ in $k-m_*$
plane for the hydrodynamic sources with different initial radii and energy
densities.  In the calculations, we take $\eta$ in the region $(-1,1)$ and
the chemical potential of boson-antiboson pairs, $\mu_{1,2}(r)=0$.
The freeze-out temperature of $\phi$ meson is taken to be 140 MeV
\cite{Padula06,Padula10,Padula10a,YZHANG14,YZHANG15}.
The variation of the BBC functions with the modified mass $m_*$ is similar
to that of the BBC functions calculated previously
\cite{Padula06,Padula10,Padula10a,YZHANG14,YZHANG15}.
However, the BBC functions for hydrodynamic sources exhibit oscillations
as a function of $k$, compared to the BBC functions for sources with
Gaussian spatial distributions and a temporal distribution of the
exponential decay \cite{Padula06,Padula10,Padula10a,YZHANG14,YZHANG15}.

\begin{figure}[htbp]
\includegraphics[scale=0.5]{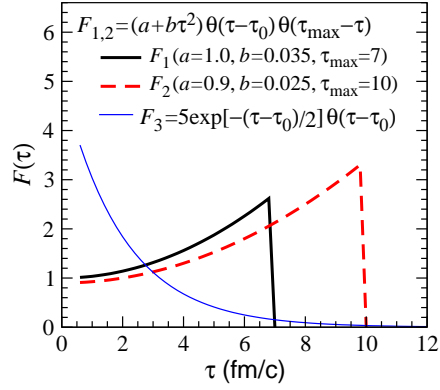}
\vspace*{-3mm}
\caption{(Color online) Parameterized temporal distributions $F_1$
and $F_2$ in Eq. (\ref{dNdtab}) for the parameter sets [$a=$ 1.0
(fm/$c$)$^{-1}$, $b=$ 0.035 (fm/$c$)$^{-3}$, $\tau_{\rm max}=7$ fm/$c$]
and [$a=$ 0.9 (fm/$c$)$^{-1}$, $b=$ 0.025 (fm/$c$)$^{-3}$, $\tau_{\rm max}
=10$ fm/$c$].  The thin solid (blue) line $F_3$ is for the temporal 
distribution of the exponential decay. }
\label{Ftau123}
\end{figure}

\begin{figure}[!htbp]
\includegraphics[scale=0.65]{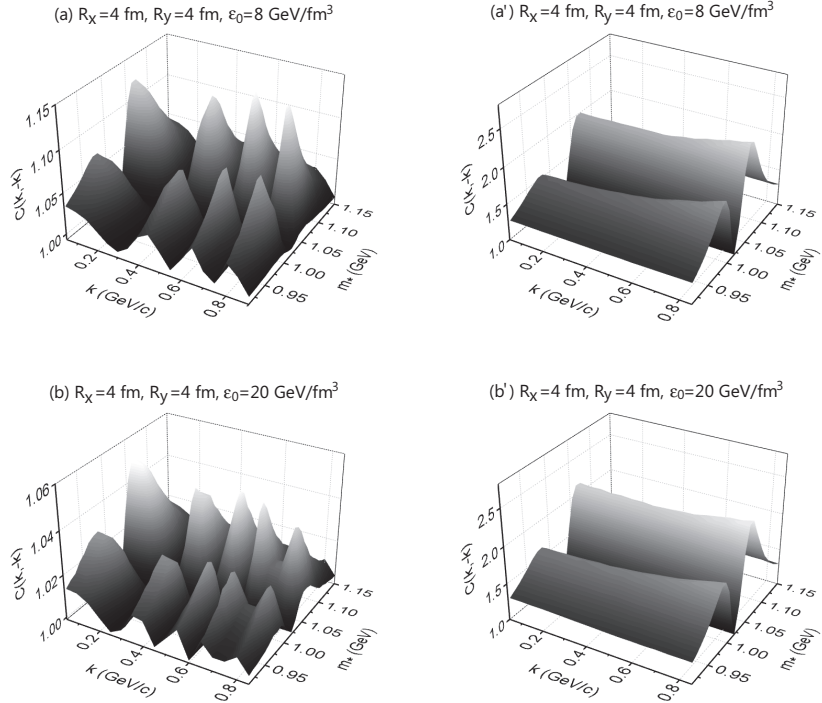}
\vspace*{-5mm}
\caption{BBC functions of $\phi\phi$ for the sources with hydrodynamic 
spatial distributions [see Fig. \ref{desrt}(b)] and parameterized temporal 
distributions (see Fig. \ref{Ftau123}).  (a, b) With temporal distributions 
$F_1$ and $F_2$ in Fig. \ref{Ftau123}; (a$'$, b$'$) with temporal distribution 
$F_3$ in Fig. \ref{Ftau123}. }
\label{BBCFp1}
\end{figure}

To examine the reason for the oscillations of BBC functions, we
calculate the BBC functions for sources with spatial distributions
obtained from the hydrodynamic freeze-out points, $dN/d|\mr_{\perp}|$,
[see Figs. \ref{desrt}(a) and \ref{desrt}(b)], and the parameterized 
temporal distribution,
\begin{equation}
\label{dNdtab}
F(\tau)=(a+b\,\tau^2)\,\theta(\tau-\tau_0)\,\theta(\tau_{\rm max}
-\tau),
\end{equation}
where $a$, $b$, and $\tau_{\rm max}$ are three parameters.  As shown in Fig.
\ref{Ftau123}, temporal distributions $F_1$ and $F_2$ are similar to the
distributions in Fig. \ref{desrt}(d).  In Fig. \ref{Ftau123}, the thin solid
line is for the temporal distribution of the exponential decay
\cite{Padula06,Padula10,Padula10a,YZHANG14,YZHANG15}.  It is much different
from the distributions of Eq. (\ref{dNdtab}).  We plot in Fig. \ref{BBCFp1}
the BBC functions of $\phi\phi$ for sources with hydrodynamic spatial
distributions [see Fig. \ref{desrt}(b)] and parameterized temporal
distributions of Eq. (\ref{dNdtab}) and exponential decay (see Fig.
\ref{Ftau123}).  Here, Figs. \ref{BBCFp1}(a) and \ref{BBCFp1}(b) are with 
temporal distributions $F_1$ and $F_2$ in Fig. \ref{Ftau123}, and Figs. 
\ref{BBCFp1}(a$'$) and \ref{BBCFp1}(b$'$) are with temporal distribution 
$F_3$ in Fig. \ref{Ftau123}.
In the calculations, the source temperature is taken to be $T_f=140$ MeV,
and the source velocities used are still the hydrodynamic source velocities
at the freeze-out points.  One can see that there are oscillations in the
BBC functions calculated with temporal distributions $F_1$ and $F_2$,
which have sharp falls to 0 at long times.  However, the oscillations
disappear in the BBC functions calculated with the temporal distribution
of exponential decay, $F_3$.  The peak values of the BBC functions for 
temporal distributions $F_1$ and $F_2$ are smaller than those of the BBC
functions for temporal distribution $F_3$.  The reason, as will be seen, 
is that the width of temporal distribution $F_3$ is much smaller than 
those of $F_1$ and $F_2$.

\begin{figure}[htbp]
\includegraphics[scale=0.5]{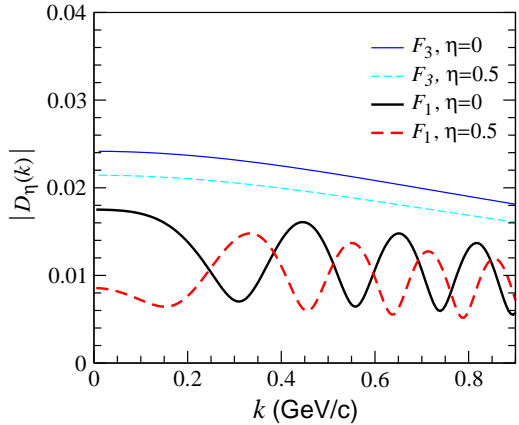}
\vspace*{-3mm}
\caption{(Color online) Magnitudes of $D_{\eta}(k)$ for temporal
distributions $F_1$ and $F_3$ in Fig. \ref{Ftau123}. }
\label{Dket}
\end{figure}

In the calculations of the BBC functions with parameterized temporal
distributions, we have
\begin{equation}
\label{BBCFD}
C(\mk,-\mk) = 1 + \frac{\left |\int_{\eta_1}^{\eta_2}\! D_{\eta}(k)
I^s_{\eta}(\mk)\,d\eta \right |^2}{\left[\int_{\eta_1}^{\eta_2}\! I^c_{\eta}(\mk)\,d\eta\right]\left[\int_{\eta_1}^{\eta_2}\! I^c_{\eta}
(-\mk)\,d\eta\right]},
\end{equation}
where
\begin{equation}
\label{Dk}
D_{\eta}(k)=\frac{\int\! F(\tau) e^{i2\omega_k\tau\cosh\eta}\,d\tau}
{|\eta_2-\eta_1|\int\! F(\tau)\,d\tau},
\end{equation}
\begin{equation}
\label{Ik}
I^s_{\eta}(\mk)=\int\!\!\frac{dN}{d^2\mr_{\perp}}\big[s^*_{\mk,-\mk}
(r)c_{-\mk,\mk}(r)n_{\mk,-\mk}(r)+c_{\mk,-\mk}(r)s^*_{-\mk,\mk}(r)
(n_{\mk,-\mk}(r)+1)\big]d^2\mr_{\perp},
\end{equation}
\begin{equation}
I^c_{\eta}(\mk)=\int \!\!\frac{dN}{d^2\mr_{\perp}}\big[|c_{\mk,\mk}
(r)|^2 n_{\mk,\mk}(r)+|s_{\mk,\mk}(r)|^2(\,n_{\mk,\mk}(r)+1)\big]
{d^2\mr_{\perp}},
\end{equation}
\begin{equation}
c_{\mk_1,\mk_2}(r)=\cosh[\,f_{\mk_1,\mk_2}(r)\,],~~~~~
s_{\mk_1,\mk_2}(r)=\sinh[\,f_{\mk_1,\mk_2}(r)\,].
\end{equation}
For the parameterized temporal distribution in Eq. (\ref{dNdtab}),
we have
\begin{equation}
|D_{\eta}(k)|=\frac{\left[b^2d^4+4a(a+bd^2)\sin^2(d\sqrt{m^2+k^2}
\cosh\eta)\right]^{1/2}}{2\sqrt{m^2+k^2}\,\cosh\eta\,(ad+bd^3/3)
|\eta_2-\eta_1|},
\end{equation}
where $d=(\tau_{\rm max}-\tau_0)$, and the approximation $b\ll a (m^2
+k^2)(\cosh\eta)^2/(c\hbar)^2$ is taken.  The oscillations of the BBC
functions are from $D_{\eta}(k)$.  We plot $|D_{\eta}(k)|$ in Fig.
\ref{Dket} for the parameterized temporal distributions in Eq.
(\ref{dNdtab}) (thick lines) and the temporal distribution of
exponential decay (thin lines), in this case  $|D_{\eta}(k)|=\left
\{(\eta_2 -\eta_1)^2[1+4(m^2\!+\!k^2)\cosh^2\!\eta\,\Delta t^2]\right
\}^{-1/2}$ and $\Delta t=2$ fm/$c$.  We take $\eta_1=-1$ and $\eta_2=1$
in the calculations.  The magnitudes of $D_{\eta}(k)$ behave as oscillations
for the parameterized temporal distributions $F_1$ because of the sharp
falls in the distributions at long times (see thick lines in Fig.
\ref{Ftau123}).  However, the magnitudes of $D_{\eta}(k)$ for the
temporal distribution of the exponential decay, $F_3$ in Fig. \ref{Ftau123},
are smoothed.  The magnitude of $D_{\eta}(k)$ for $F_3$ are larger than
those for $F_1$ because the width of the $F_3$ distribution is smaller
than that of $F_1$.  For all the temporal distributions, the magnitude
of $D_{\eta}(k)$ decreases with increasing $\eta$.  And, the BBC functions
are determined by the product of $D_{\eta}(k)$ and $I^s_{\eta}(k)$.

\vspace*{5mm}

\subsection{Dependence of the BBC function on the direction of the 
particle momentum}
\begin{figure}[htbp]
\includegraphics[scale=0.65]{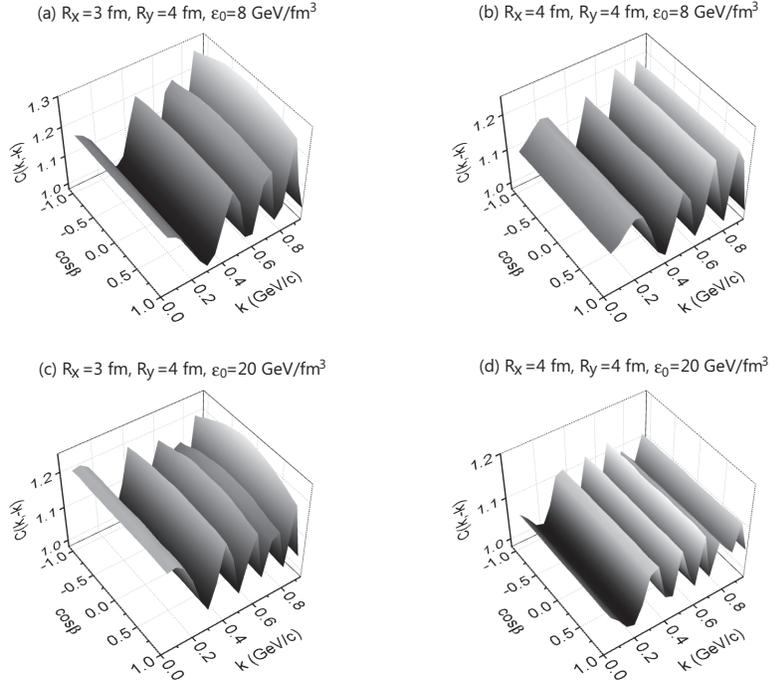}
\vspace*{-5mm}
\caption{BBC functions of $\phi\phi$ in the $\cos\beta$-$k$ plane
for hydrodynamic sources with different initial radii and energy
densities.  Here, $m_*$ is taken as 1.05 GeV. }
\label{BBCFpb}
\end{figure}

For anisotropic sources, the BBC functions depend not only on the
magnitude of the particle momentum, but also on its direction
\cite{YZHANG15}.  This is because the source velocities in different
directions lead to the variation of $k^{\mu}u_{\mu}$ with the momentum
direction.  
To examine the dependence of the BBC functions on the direction of the
particle momentum, we use
\begin{equation}
\cos\alpha=k_z/|{\mk}|,~~~~~\cos\beta=k_x/|\mk_T|,~~~~
\left(|\mk_T|=\sqrt{k_x^2+k_y^2}\,\right),
\end{equation}
to describe the direction of the particle momentum.  $\alpha$ and $\beta$
are the polar angle and azimuthal angle of the particle.  In Fig.
\ref{BBCFpb}, we plot the BBC functions of $\phi\phi$ in the $\cos\beta$-$k$
plane for hydrodynamic sources with different initial radii and energy
densities.  Here, $m_{\!*}$ is taken as 1.05 GeV, corresponding to
approximately the peaks of the BBC functions (see Fig. \ref{BBCFP34}).
For $R_x<R_y$, the source expanding velocity in the $x$ direction is higher
than that in the $y$ direction.  The BBC functions for sources with
$R_x<R_y$ are larger at $\cos\beta=0$ than those at $\cos\beta=1$,
because the average values of $e^{-k^{\mu}u_{\mu} /T_f}$ are smaller
at $\cos\beta=0$ than that at $\cos\beta=1$ in this case \cite{YZHANG15}.
For the source with $R_x=R_y$, the BBC functions are independent of
$\cos\beta$.

\begin{figure}[!htbp]
\vspace*{5mm}
\includegraphics[scale=0.6]{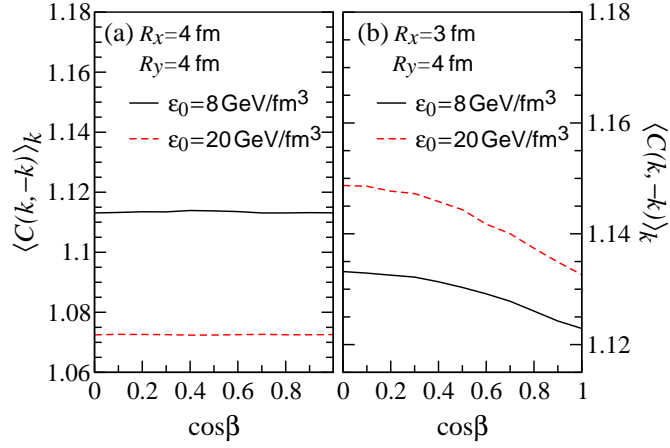}
\vspace*{-5mm}
\caption{(Color online) Dependence of the average BBC functions,
$\langle C(k,-k)\rangle_k$, of $\phi\phi$ on the cosine of the 
particle azimuthal angle for hydrodynamic sources with different 
initial radii and energy densities. Here, $m_*$ is taken as 1.05 
GeV and the momentum region averaged is 0--1 ${\rm GeV}\!/\!c$. }
\label{BBCFPam}
\end{figure}

In Fig. \ref{BBCFPam}, we show the dependence of the average BBC functions,
$\langle C(k,-k)\rangle_k$, of $\phi\phi$ on the cosine of the particle
azimuthal angle for hydrodynamic sources with different initial radii
and energy densities.  Here, $m_*$ is taken as 1.05 GeV and the momentum
region averaged is 0--1 ${\rm GeV}\!/\!c$.  The BBC functions are
independent of the azimuthal angle for transverse isotropic sources.
However, the BBC functions for transverse anisotropic sources increase
with increasing azimuthal angle of the particles ($0<\beta<\pi/2$).  
For transverse isotropic sources, the average BBC function for the higher
initial energy density is smaller than that for the lower initial energy
density.  The reasons are that the source with a higher initial energy
density has a larger expansion velocity and a wider temporal distribution
of freeze-out points.  For transverse anisotropic sources, we observe that 
the average BBC function for the higher initial energy density is larger
than that for the lower initial energy density.  This is mainly because 
the many oscillations of the BBC function for the source with the higher 
initial energy density lead to an increase in the average value in the 
momentum region [see Figs. \ref{BBCFpb}(a) and \ref{BBCFpb}(c)].

\begin{figure}[htbp]
\includegraphics[scale=0.65]{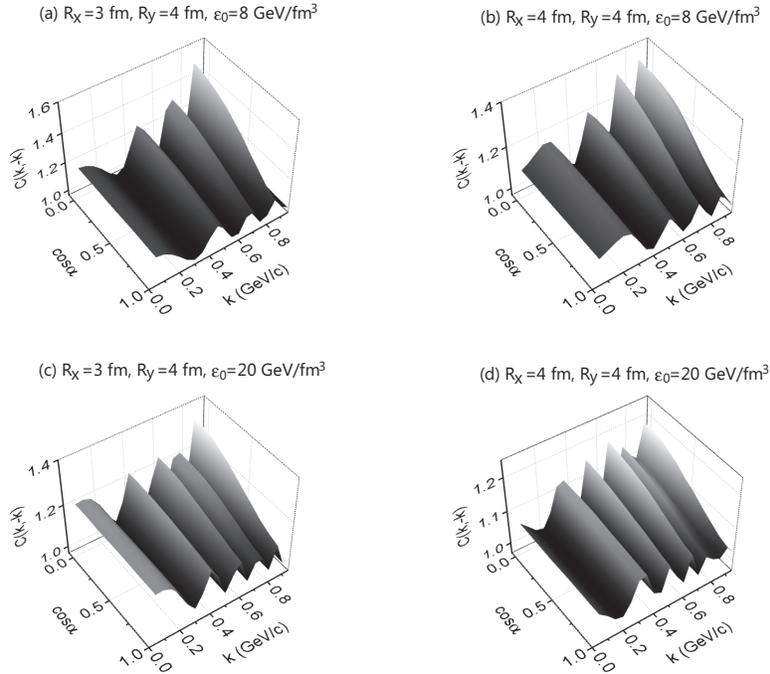}
\vspace*{-5mm}
\caption{BBC functions of $\phi\phi$ in the $\cos\alpha$-$k$ plane
for hydrodynamic sources with different initial radii and energy
densities.  Here, $m_*$ is taken as 1.05 GeV. }
\label{BBCFpa}
\end{figure}

We plot in Fig. \ref{BBCFpa} the BBC functions of $\phi\phi$ in the 
$\cos\alpha$-$k$ plane for hydrodynamic sources with different initial
radii and energy densities, and for $m_{\!*}=1.05$ GeV.  The ridge
values of the BBC functions for a fixed $k$ decrease with increasing
$\cos\alpha$.  The reasons for this are that the average source 
longitudinal velocity is higher than the average source transverse 
velocity for hydrodynamic sources with Bjorken longitudinal boost 
invariance \cite{Bjo83}, and the higher longitudinal velocity leads 
to smaller average values of $e^{-k^{\mu}u_{\mu}/T_f}$ at $\cos\alpha=0$
than at $\cos\alpha=1$ \cite{YZHANG15}.

\begin{figure}[!htbp]
\vspace*{5mm}
\includegraphics[scale=0.6]{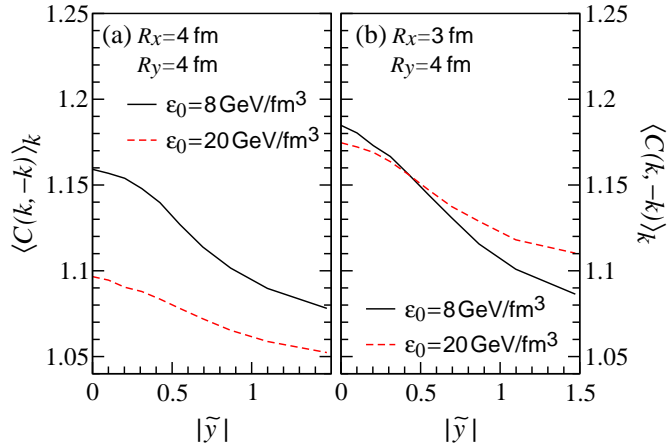}
\vspace*{-5mm}
\caption{(Color online) Dependence of the average BBC functions
$\langle C(k,-k) \rangle_k$ of $\phi\phi$ on the pseudorapidity of 
the particle for hydrodynamic sources with different initial radii 
and energy densities.  Here, $m_*$ is taken as 1.05 GeV and the 
momentum region averaged is 0--1 ${\rm GeV}\!/\!c$. }
\label{BBCFPy}
\end{figure}

Because $\cos\alpha$ is related to particle pseudorapidity by
\begin{equation}
{\widetilde y}=\tanh^{-1}(\cos\alpha),
\end{equation}
the polar angle dependence of the BBC functions can lead to 
pseudorapidity dependence of the BBC functions.  In Fig. \ref{BBCFPy},
we show the dependence of the average BBC functions, $\langle C(k,-k)
\rangle_k$, of $\phi\phi$ on the pseudorapidity of the particle for
hydrodynamic sources with different initial radii and energy densities.
Here, $m_*$ is taken as 1.05 GeV and the momentum region averaged is 0--1
${\rm GeV}\!/\!c$.  The BBC functions decrease with increasing $|\widetilde
y|$ as expected.  Because the average transverse velocities for sources
with a higher initial energy density are higher than those for sources
with a lower initial energy density, the BBC functions for sources with
a higher initial energy density decrease more slowly with increasing 
$|\widetilde y|$.

\section{Summary and discussion}
In the hot and dense hadronic sources formed in high-energy 
heavy-ion collisions, particle interactions in medium might lead 
to a squeezed BBC of boson-antiboson pairs.  The investigations 
of the BBC in previous works are for sources with space-time-separated 
source distributions
\cite{Padula06,Padula10,Padula10a,YZHANG14,YZHANG15}.  The smoothed
temporal distribution of the exponential decay scales the BBC functions
and leads to monotonic BBC functions with respect to the particle
momentum \cite{Padula06,Padula10,Padula10a,YZHANG14,YZHANG15}.

Relativistic hydrodynamics is successful in describing the source
space-time evolution in high-energy heavy-ion collisions.  In this
paper, we investigate the BBC functions of $\phi\phi$ for sources
evolving hydrodynamically in ($2+1$) dimensions and with longitudinal
boost invariance.  For hydrodynamic sources, the BBC functions
oscillate as a function of the particle momentum.  The reason for 
the oscillations is that the temporal distributions of sources
evolving hydrodynamically have sharp falls to 0 at large evolving
times, compared to the temporal distribution of the exponential 
decay.  We also investigate the dependences of the BBC functions on 
the directions of the particle momentum.  For transverse anisotropic
sources, the anisotropic source velocity leads to the dependence of
the BBC functions on the particle azimuthal angle.  The BBC functions
are minimum when the azimuthal angles of the particles reach 0.  
Because the average source longitudinal velocity is higher than the
average source transverse velocity, the BBC functions increase with
decreasing absolute value of the particle pseudorapidity.
The oscillations and the dependences on the particle azimuthal angle
and pseudorapidity are the significant signatures for detecting
the BBC in high-energy heavy-ion collisions.

For $\mk_1=\mk$, $\mk_2=-\mk_1=-\mk$, $e^{2iK_{1,2}\cdot r}=e^{2i
\omega_{\mk}t}$, the BBC function $C(\mk,-\mk)$ for the hydrodynamic
source is related to the temporal Fourier transformation of the
space-time distribution of source freeze-out points [see Eqs.
(\ref{BBCf}), (\ref{Gchydro}), and (\ref{Gshydro})].  So, the BBC
function is very sensitive to the temporal distribution of the source,
and an appropriate source space-time distribution is important for
estimating the BBC effect in high-energy-heavy ion collisions.  In
Refs. \cite{AsaCsoGyu99,Padula06,Padula10,Padula10a,YZHANG14,YZHANG15},
a sudden freeze-out assumption at time $\tau_f$ is adopted, and then
a parameterized distribution of $\tau_f$ (exponential decay) is used
in the calculations to suppress the BBC functions.
In Ref. \cite{Kno11}, the author argues the appropriateness of the
sudden freeze-out assumption of time and discusses the exponential
suppression of the BBC function based on three parameterized
temporal distributions.  In this work, we extract the space-time
distributions of the source from the ($2+1$)-dimensional hydrodynamics.
The temporal and spatial distributions of the source are related in
the calculations.  Although it is a significant advance compared
to the parameterized space-independent distributions of time used in
\cite{AsaCsoGyu99,Padula06,Padula10,Padula10a,YZHANG14,YZHANG15,Kno11},
further investigations based on more realistic models in which
the model parameters are determined by the experimental data on the
observables, such as single particle spectra, elliptic flow, HBT
radii, are needed for the expectations of the BBC effect in 
high-energy heavy-ion collisions.

\begin{acknowledgments}
This research was supported by the National Natural Science Foundation
of China under Grant No. 11275037.
\end{acknowledgments}

\end{document}